\documentstyle[aps,epsf]{revtex}

\begin{document}

\title{Critical couplings for chiral symmetry breaking via instantons}

\author{F. S. Roux\\ {\it Department of Physics, University
of Toronto \\ Toronto M5S1A7, CANADA } }

\maketitle

\begin{abstract} 
Using an instanton effective action formalism, we compute the critical coupling for
the nonperturbative formation of a dynamical mass via instantons in non-Abelian gauge
theories with $N_f$ massless fermions. Only continuous phase transitions are
considered. For large values of $N_f$ the critical couplings are found to be much
smaller than the equivalent critical couplings obtained from gauge exchange
calculations in the ladder approximation.
\end{abstract}


\section{introduction}
\label{intro}

Nonperturbative gauge dynamics present many challenges that are yet to be fully
understood. One of these is the phenomenon of chiral symmetry breaking. A theory
with an $SU(N_c)$ gauge symmetry and $N_f$ fermions in the fundamental representation
has a global chiral symmetry in the massless limit. Provided that $N_f < 5.5 N_c$
this theory is asymptotically free. Close to this threshold the coupling is bounded
above due to an infrared fixed point in the beta function. As $N_f$ becomes smaller
the fixed point value increases. Below a critical value for $N_f$ the coupling can
grow strong enough to break the chiral symmetry down to the vector symmetry:
\begin{equation}
SU(N_f)_R \times SU(N_f)_L \times U(1)_V \rightarrow SU(N_f)_V \times U(1)_V .
\label{v1}
\end{equation}
As a result the fermions acquire a dynamical mass, which acts as an order parameter
for the spontaneous breaking of the chiral symmetry. This much is well understood,
but the respective parts played by gauge exchanges and instantons in this process
still needs clarification. 

It has been known for some time that gauge exchanges can generate a dynamical
mass. To leading order (ladder approximation) in Landau gauge the critical coupling
associated with gauge exchange dynamics is
\begin{equation}
\alpha_c^{ge} = {2\pi N_c \over 3 (N_c^2-1)} .
\label{v2}
\end{equation}
Assuming a two-loop beta function, one can show that the critical number of flavors,
below which dynamical chiral symmetry breaking occurs, is \cite{r_atw}
\begin{equation}
N_f^c = 4 N_c - {6 N_c \over 25 N_c^2-15} \approx 4 N_c .
\label{v3}
\end{equation}

It is known that instanton \cite{r_bpst} dynamics can also break the chiral symmetry
\cite{r_c,r_cc}. Attempts to investigate this possibility have been hampered by the
presence of an infrared divergence in the integral over the instanton size. In their
recent investigation of the critical number of flavors for chiral symmetry breaking
via instantons, Appelquist and Selipsky \cite{r_as} avoided the problem of infrared
divergences by introducing the following two modifications of the Carlitz and
Creamer gap equation \cite{r_cc}:
\begin{itemize}
\item They introduced a mass dependence for the fermion determinant that is not only
valid for the small mass limit \cite{r_t} but also for the large mass limit
\cite{r_ag}. This function simply makes a sharp transition from the small mass
behavior to the large mass behavior. The effect is to introduce a fermion mass scale
below which the fermions are integrated out.
\item In Reference \cite{r_as} all couplings in the instanton expression are allowed
to run according to the two-loop beta function. For those values of $N_f$ where the
beta function has a fixed point the coupling remains fairly constant until it
reaches the fermion mass scale. Fermions are integrated out below the fermion mass
scale. This causes the beta function to revert back to normal asymptotic free
running, which gives a suppression of the integrand in the infrared region because
of the way it depends on the coupling.
\end{itemize}
The result of their investigation was that the critical number of flavors for
instantons is about $4.77 N_c$. From this they concluded that the instanton
contribution to chiral symmetry breaking is comparable to that of gauge exchanges.

\begin{figure} 
\centerline{\epsfysize = 6 cm \epsfbox{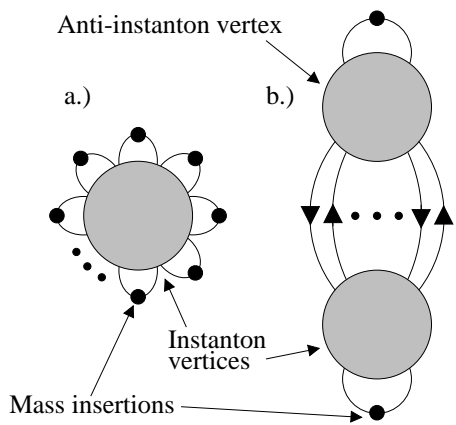}}
\caption{Leading instanton diagrams: a) one--instanton, b)
instanton--anti-instanton with two mass insertions.}
\label{diagramme}
\end{figure}

In this paper we reinvestigate chiral symmetry breaking through instantons, but
our investigation differs from that of Reference \cite{r_as} in two essential
aspects:
\begin{itemize}
\item The recently proposed instanton effective action formalism \cite{r_rt} is used
to determine the conditions under which the symmetric vacuum becomes unstable. This
formalism reproduces the Carlitz and Creamer gap equation in the small mass limit
for $SU(2)$. However, we are not interested in the shape of the mass function, but
rather in the critical couplings and the critical number of flavors.
\item We concentrate on continuous phase transitions, such as found for chiral
symmetry breaking through gauge exchanges. The effective action formalism makes it
clear that the instanton contribution responsible for this type of transition is not
the one--instanton amplitude of Figure \ref{diagramme}a but rather the
instanton--anti-instanton amplitude with two mass insertions, shown in Figure
\ref{diagramme}b.
\end{itemize}

The latter point needs clarification. The effective action can be interpreted as an
effective potential. This makes it easier to determine when the stability of the
symmetric vacuum is affected by the dynamics. For small couplings the global minimum
of the effective potential is located at the origin, where the order parameter is
zero. When the coupling is increased the global minimum remains at the origin until
the coupling crosses the critical value where the phase transition occurs. At this
point the global minimum can either jump discontinuously to a nonzero value of the
order parameter, indicating a first order phase transition, or it can move
continuously (though non-analytically) to a nonzero value of the order parameter,
indicating a continuous phase transition. For a first order phase transition a local
minimum must develop as the coupling is increased. This becomes deeper as the
coupling is increased further until it is below the minimum at zero and so becomes
the global minimum when the coupling crosses the critical value. A continuous phase
transition appears when the global minimum `rolls' out of the origin.  This implies
that the minimum at the origin must flatten out and become unstable. To investigate
first order phase transitions one must be able to analyze the effective potential at
arbitrarily large values of the order parameter. Unfortunately the effective
potential is not reliable at large values of the order parameter because it is not
possible to make a reliable truncation of the series of diagrams in the effective
potential for large order parameters. For a continuous phase transition, on the
other hand, the order parameter would be arbitrarily small close enough to the
transition point. Hence, the order parameter can be used as an expansion parameter,
allowing one to make reliable truncations of the series of diagrams. For this reason
we restrict ourselves to the investigation of continuous phase transitions.

In a stability analysis for continuous phase transitions the one--instanton
amplitude is irrelevant. The reason is as follows. The effective potential consists
of terms with positive powers of the order parameter.  Near the origin the
nondynamical kinetic part of the potential is second order in the order parameter.
Hence, to destabilize the potential at the origin the dynamical terms must be of
second order or smaller.  Terms with higher powers of the order parameter will
become insignificant compared to the kinetic term when the order parameter
approaches zero. Looking at the one--instanton diagram, shown in Figure
\ref{diagramme}a, one sees that this term always comes with as many powers of the
order parameter as it takes to close off all the zeromode lines. For $N_f$ flavors
there would be $N_f$ factors of the order parameter. This means that for $N_f > 2$
the one--instanton term cannot compete with the kinetic term near the origin. The
leading contribution to the dynamic part of the potential that is second order in
the order parameter is the instanton--anti-instanton term with two mass insertions
shown in Figure \ref{diagramme}b.  Therefore only this term is considered in our
calculation. It turns out that this term does not have an infrared problem in our
stability analysis.

The main result of our stability analysis is presented in terms of an expression for
arbitrary $N_f$ and $N_c$ from which one can compute the critical couplings for
chiral symmetry breaking via instantons. We determine the critical numbers of flavors
for $N_c = 3, 4, 5$ and $6$, assuming a two-loop beta function. Numerical values for
critical couplings are provided in the fixed point regions of these values of $N_c$,
up to the respective critical numbers of flavors. These results indicate that the
critical couplings for the formation of 2-point functions via instantons are much
smaller than those for gauge exchange dynamics when the number of flavors becomes
large. The increasing strength of the instanton dynamics can be understood in terms
of the increase in the number of fermion lines of the instanton vertices.

The instanton effective action formalism is briefly reviewed in Section
\ref{s_diag}. We distinguish between the dynamical and nondynamical terms of the
effective action. Section \ref{nondyn} addresses the nondynamical terms and the
dynamical term is considered in Section \ref{dyn}. We present and discussed the
results of this calculation in Section \ref{results} and conclude with a short
summary in Section \ref{summary}.

\section{Instanton effective action formalism}
\label{s_diag}

The 2-point effective action \cite{r_dm,r_cjt} in the instanton formalism of
Reference \cite{r_rt} is given by
\begin{equation}
\Gamma[S] = {\rm Tr} \left\{ \ln \left( S^{-1} \right) \right\} + {\rm Tr}
\left\{ \left( S_m^{-1} - S^{-1} \right) S \right\} + W_{2PI}[S]
\label{v4}
\end{equation}
where the fermion {\it modal propagator}, $S_m$, consists of different propagators
for the different types of modes associated with an instanton background
configuration. We parameterize the full fermion propagator $S$ by
\begin{equation}
S(p) = {1 \over i p\!\!\!/ - \Sigma(p)} = {- i p\!\!\!/ - \Sigma(p) \over p^2 +
\Sigma^2(p)}
\label{v5}
\end{equation}
where $\Sigma(p)$ is the dynamical mass function.

One can classify the terms of the effective action as either {\it dynamical} or {\it
nondynamical}. Dynamical terms are those that vanish when the coupling is set to
zero, while the nondynamical terms are those that remain. The first two one-loop
kinetic terms in (\ref{v4}) are nondynamical terms. They consist of one-loop diagrams
that are only composed of the full fermion propagator $S$ and the fermion modal
propagator, $S_m$.

The dynamical term $W_{2PI}[S]$ in (\ref{v4}) is the sum of all 2PI vacuum diagrams
constructed with Feynman rules for:
\begin{itemize} 
\item the $2 N_f$-point instanton vertices, ${\cal V}_I$ and ${\cal V}_A$, given
below;
\item the full fermion propagator $S$, given in (\ref{v5}); and
\item an integral over the collective coordinates (sizes, positions and color
orientations) of all instantons in the diagram.
\end{itemize}

The dynamics in this analysis are provided by the instantons and are represented by
the instanton vertices:
\begin{equation}
{\cal V}_I = {\cal A} \prod_n^{N_f} \left( \delta_0 S_0^{-1} \psi^0_n \right) \left(
\overline{\psi}^0_n S_0^{-1} \overline{\delta}_0 \right) ,
\label{v7}
\end{equation}
where $S_0^{-1}$ is the inverse bare propagator and the functional derivatives
$\delta_0 (= \delta/ \delta \eta_0)$ and $\overline{\delta}_0 (= \delta/ \delta
\overline{\eta}_0)$ operate on the zeromode parts of the fermion sources, $\eta_0$
and $\overline{\eta}_0$. The expression for anti-instantons, ${\cal V}_A$, differs
from the expression in (\ref{v7}) only in that the zeromode functions, $\psi^0$ and
$\overline{\psi}^0$, have the opposite helicities. 

The coefficient of the instanton vertex is the integrand of the 't Hooft instanton
amplitude \cite{r_t}:
\begin{equation}
{\cal A} = \kappa \gamma(\alpha) \frac{1}{\rho^5} (\mu\rho)^{b_0+N_f} .
\label{v8}
\end{equation}
Here $\rho$ is a scale parameter for the size of the instanton and $b_0 =
\frac{1}{3}(11 N_c - 2 N_f)$.

The dependence on the gauge coupling in (\ref{v8}) is contained in
\begin{equation}
\gamma \left( \alpha(\mu) \right) = \left( {2 \pi \over \alpha(\mu)} \right)^{2
N_c} \exp \left(- {2 \pi \over \alpha(\mu)} \right) .
\label{v9}
\end{equation}
The factor $(\mu\rho)^{b_0}$ appearing in (\ref{v8}) can be incorporated into
the exponent part of $\gamma(\alpha)$. This then leads to one-loop running as a
function of $\rho$ for the coupling in the exponent in $\gamma(\alpha)$. The
instanton expression in (\ref{v8}) can be computed to higher orders which would
cause the coupling in the monomial part of $\gamma(\alpha)$ to run as a function of
$\rho$ as well, but always at one order less than the coupling in the exponent. To
all orders one would expect that both couplings run according to the exact beta
function. Hence, assuming that the two-loop beta function is an accurate enough
representation of the exact beta function, one can allow both couplings in
$\gamma(\alpha)$ to run according to the two-loop beta function \cite{r_as}. This is
what we shall assume so that we can make the replacement $\gamma(\alpha(\mu))
(\mu\rho)^{b_0} \rightarrow \gamma(\alpha(\rho))$.

\begin{figure} 
\centerline{\epsfysize = 8 cm \epsfbox{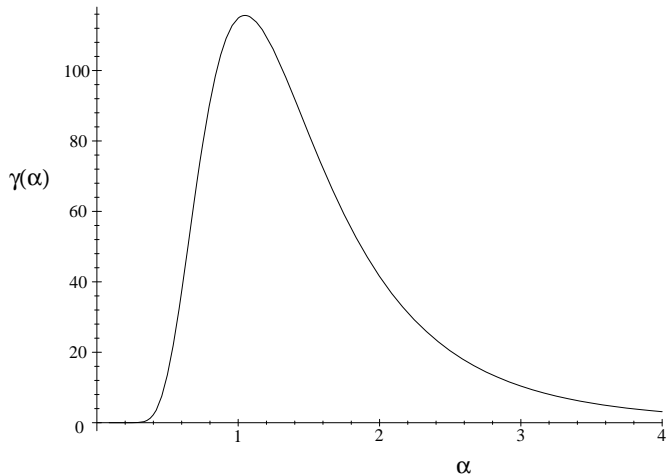}}
\caption{The shape of $\gamma(\alpha)$ for $N_c=3$.}
\label{gamma}
\end{figure}

The shape of $\gamma(\alpha)$ as a function of $\alpha$ is shown in Figure
\ref{gamma}. The instanton amplitude peaks at a value of $\alpha = \pi/N_c$.

The prefactor in (\ref{v8}) is
\begin{equation}
\kappa = {2 \exp(5/6) \exp(B N_f-C N_c]) \over \pi^2 (N_c-1)! (N_c-2)!}
\label{v10}
\end{equation}
where we follow \cite{r_as} and use the one-loop values: $B=0.2917$, $C=1.5114$ in
MS-bar \cite{r_l}. 

Note that unlike the case in Reference \cite{r_rt} the integration over the color
orientations has not yet been done in (\ref{v7}).  To alleviate the combinatorical
calculations, this step is postponed until after the propagators are connected and
unnecessary gauge transformations are cancelled. However, $\kappa$ in
(\ref{v10}) already contains the volume of the factor group, $SU(N_c)/G$, where $G$
is the subgroup that leaves the instantons invariant.  All that remains of the
integration over color orientations is a color averaging integral.

The expression of the fermion zeromode function in singular gauge that appears in
(\ref{v7}) is
\begin{equation}
\psi^0(x) = {\sqrt{2} \rho\ x_{\mu} \gamma_{\mu} U^{\alpha} \over \pi \sqrt{\mu}
\sqrt{x^2} (\rho^2+x^2)^{3/2}} ,
\label{v11}
\end{equation}
where $\mu$ is the renormalization scale\footnote{The $\mu$ is included here to get
a mass dimension of $\frac{3}{2}$ for the fermion zeromodes.} and $x^2$ denotes an
Euclidean dot-product: $x_{\mu} x_{\mu}$. The Euclidean Dirac matrices,
$\gamma_{\mu}$, are here defined so that $\{ \gamma_{\mu}, \gamma_{\nu} \} = 2
\delta_{\mu\nu} \cdot {\bf 1}$, ${\rm tr} \{ \gamma_{\mu} \gamma_{\nu} \} = 4
\delta_{\mu\nu}$ and $\gamma_{\mu}^{\dag} = \gamma_{\mu}$. The $U^{\alpha}$ denotes
a Dirac spinor with isospin index $\alpha$. (The isospin index is implicit in
$\psi^0(x)$.) This spinor is normalized such that $\sum_{\alpha}
\overline{U}_{\alpha} U^{\alpha} = 1$ and it obeys the identity
$\sum_{\alpha} U^{\alpha} \overline{U}_{\alpha} = \frac{1}{4} (1+\gamma_5)$.

We shall also later need the Fourier-transform of the zeromode function
\begin{equation}
\Psi^0(p) = \int \psi^0(x) \exp(i x \cdot p)\ d^4 x = i {2 \pi \sqrt{2} \rho\ p_{\mu}
\gamma_{\mu} U^{\alpha} \over \sqrt{\mu} p^2} f\left( \frac{\rho \sqrt{p^2}}{2}
\right) ,
\label{v12}
\end{equation}
where the function,
\begin{equation}
f(x) = 2 x \left[ I_0(x) K_1(x) - I_1(x) K_0(x) \right] - 2 I_1(x) K_1(x) ,
\label{v13}
\end{equation}
is defined in terms of modified Bessel functions \cite{r_AS}.

\section{The nondynamical terms}
\label{nondyn}

The trace over the inverse modal propagator in (\ref{v4}) contains an integral over
all the collective coordinates involved in the diagram. The general form of the
result of this integration is the inverse bare propagator, $S_0^{-1} = i p\!\!\!/$,
multiplied by a momentum-dependent scalar form factor, which we shall assume to be
equal to 1. 

In terms of (\ref{v5}) the nondynamical terms in (\ref{v4}), which we denote by
$\Gamma_{kin}$, becomes
\begin{equation}
\Gamma_{kin} = - \int \left( {4 \Sigma^2(p) \over p^2 + \Sigma^2(p)} - 2 \ln \left[
p^2 + \Sigma^2(p) \right] \right) {d^4p \over (2\pi)^4}.
\label{v14}
\end{equation}

It is customary to introduce an infrared cutoff at the momentum scale where $p_0 =
\Sigma(p_0)$. Close to the phase transition point $\Sigma(p)$ is very small so that
$p_0
\approx 0$. An expansion of $\Gamma_{kin}$ in term of $\Sigma(p)/p$ contains only
even powers. We drop the terms that are independent of $\Sigma(p)$ because they do
not affect the location of the minimum and we drop the terms that are higher than
second order in
$\Sigma(p)$ because they are suppressed. The result is
\begin{equation}
\Gamma_{kin}^{(2)} = - {N_c N_f \over 4 \pi^2} \int_0^{\infty} \Sigma(p)^2 p\
dp.
\label{v15}
\end{equation}

Now we make a redefinition of the integration variables and do a Fourier
transformation. For this purpose we define 
\begin{equation}
p = \mu \exp(t) ~~~ {\rm and} ~~~ \Sigma(p) = \mu \exp(-t) \int_{-\infty}^{\infty}
\sigma(\omega) \exp(i\omega t)\ d \omega
\label{w1}
\end{equation}
where $\mu$ is the renormalization scale. Now we have
\begin{eqnarray}
\Gamma_{kin}^{(2)} & = & - {N_c N_f \mu^4 \over 4 \pi^2} \int_{-\infty}^{\infty}
\int_{-\infty}^{\infty} \sigma(\omega_1) \sigma(\omega_2) \int_{-\infty}^{\infty}
\exp(i\omega_1 t) \exp(i\omega_2 t)\ dt\ d \omega_1\ d \omega_2 \nonumber \\ & = &
- {N_c N_f \mu^4 \over 2 \pi} \int_{-\infty}^{\infty} \sigma(\omega) \sigma(-\omega)\
d \omega.
\label{w2}
\end{eqnarray}

\section{The dynamical term}
\label{dyn}

The instanton--anti-instanton term that is quadratic in the dynamical mass looks as
follows:\ each mass insertion closes off two lines of an instanton vertex. The
remaining lines are all connected between the two vertices. (See Figure
\ref{diagramme}b.) The resulting diagram is expressed as
\begin{equation}
W_{2PI}^{(2)} = {\cal F} \kappa^2 \int \int \gamma(\alpha(\rho_1))\
\gamma(\alpha(\rho_2))\ (\mu^2\rho_1\rho_2)^{N_f} D(\rho_1) D(\rho_2)
G(\rho_1,\rho_2) \frac{1}{\rho_1^5} \frac{1}{\rho_2^5}\ d\rho_1\ d\rho_2 
\label{v16}
\end{equation}
where $\rho_1$ and $\rho_2$ denote the sizes of the two instantons; the
combinatorial factor ${\cal F}$ simply counts the number of terms associated with
this diagram; $D(\rho_1)$ and $D(\rho_2)$ are the parts of the expression due to the
two mass loops and $G(\rho_1,\rho_2)$ is the part of the expression due to the
remaining fermion loops.

\subsection{The combinatorics and color averaging}

To count the number of terms we start by considering all the possible ways that
the propagators can close off the lines of the two instanton vertices to give a
diagram with two mass loops. Then we perform the color averaging and add up all the
different terms. 

Recall that each line of an instanton vertex has a specific flavor associated with
it. There is one incoming and one outgoing line for each flavor. Because the
propagator is flavor diagonal it must connect lines of the same flavor. Hence, there
are $N_f$ distinct ways to connect one of the two mass insertions, each leaving the
other mass insertion and the remaining fermion lines with only one way to be
connected.

Next we perform the averaging over the color orientations of the two respective
instantons. The color orientations are represented by gauge transformations that
operate on the zeromodes:
\begin{equation}
\psi^0 \rightarrow g \psi^0 ~~~ {\rm and} ~~~ \overline{\psi}^0 \rightarrow
\overline{\psi}^0 g^{-1} .
\label{v17}
\end{equation}
In the mass loops they cancel because there we have a $g$ and a $g^{-1}$ from the
same instanton. The gauge transformations of the remaining lines can be combined
into relative gauge transformations of one instanton with respect to the other:
$g_1^{-1} g_2 = g_r$ or $g_2^{-1} g_1 = g_r^{-1}$. Thus the two color averaging
integrals become one trivial ($=1$) and one nontrivial averaging integral. The
nontrivial integral contains $(N_f-1)$ factors of $g_r$ and $g_r^{-1}$. The result
of the averaging is a series of tensor structures each multiplied by a coefficient
that depends on $N_f$ and $N_c$. General expressions for these coefficients are
provided in Reference \cite{r_sam}. The tensor structures indicate how color indices
of the same instanton are interconnected. This in turn determines how the fermion
lines between the two instantons are closed off into fermion loops and thereby
determines the form of the momentum integrals.
 
The evaluation of all these momentum integrals is too challenging to attempt.
However, we expect that the sizes of the momentum integrals do not differ by more
that an $O(1)$ factor and that the variations would average out. Therefore we set
them all equal and estimate their typical size in Section \ref{glusse}. Each
tensor structure consists of
$(N_f-1)!$ terms because of all the possible permutations of the $(N_f-1)$ pairs of
fermion lines between the two instantons. The different tensor structures denote
different {\it relative} permutations of one line with respect to the other in the
$(N_f-1)$ pairs of fermion lines. All tensor structures that are associated with the
same conjugacy class in the permutation group have the same coefficient.
 
Because we assume that the momentum integrals are the same, we only have to add all
the coefficients of all the tensor structures multiplied by the number of terms
associated with each tensor structure. It can be shown that for permutations of
$(N_f-1)$ elements,
\begin{equation}
\sum_{classes} \alpha_i N_i = {(N_c-1)! \over (N_c+N_f-2)!} ,
\label{v18}
\end{equation}
where $N_i$ is the number of group elements in the $i$-th conjugacy class of the
permutation group and $\alpha_i$ is the coefficient of the tensor structures in that
class. To get the total count we have to multiply (\ref{v18}) by $N_f$ for the number
of ways to connect the mass loops and by $(N_f-1)!$ for the number of terms in each
tensor structure. The resulting combinatorial factor is
\begin{equation}
{\cal F} = {N_f! (N_c-1)! \over (N_c+N_f-2)!} .
\label{v19}
\end{equation}

\subsection{The mass insertions}

Using (\ref{v5}) and (\ref{v12}) we obtain the following expressions for the mass
loops
\begin{eqnarray}
D(\rho) & = & \int \overline{\Psi}^0(p) S_0^{-1}(p) S(p) S_0^{-1}(p) \Psi^0(p)
{d^4p \over (2\pi)^4} \nonumber \\ & = & {\rho^2 \over \mu} \int
{\Sigma(p) \over p^2 + \Sigma^2(p)} f^2 \left( {\rho p \over 2} \right) p^3\ dp
\nonumber \\ & = & {\rho^2 \over \mu} \int \Sigma(p) f^2 \left( {\rho p \over 2}
\right) p\ dp + O \left( \Sigma^3(p) / p^3 \right)
\label{v20}
\end{eqnarray}
where we again made use of the approximations introduced in Section \ref{nondyn}.
Next we again make a redefinition of the integration variables and do a Fourier
transformation as in (\ref{w1}). In addition we define
\begin{equation}
{\rho \mu \over 2} = \exp(\eta) ~~~ {\rm and} ~~~ f \left( \exp(a) \right) = f_0(a)
\label{v21}
\end{equation}

 As a result we have
\begin{eqnarray}
D(\rho) & = & 4 \exp(2\eta) \int_{-\infty}^{\infty} \int_{-\infty}^{\infty}
\sigma(\omega) \exp(i\omega t) f_0^2(t+\eta) \exp(t)\ dt\ d \omega \nonumber \\
& = & 4 \exp(\eta) \int_{-\infty}^{\infty} \sigma(\omega) \exp(-i\omega
\eta) F(\omega)\ d \omega
\label{v22}
\end{eqnarray}
where
\begin{equation}
F(\omega) = \int_{-\infty}^{\infty} \exp(i\omega a) f_0^2(a) \exp(a)\ da .
\label{v23}
\end{equation}

\subsection{The remaining fermion loops}
\label{glusse}

The rest of the fermion lines between the two instanton vertices combine into
\begin{equation}
G(\rho_1,\rho_2) = \prod_i^{2 N_f - 2} \left[ \int {d^4p_i \over (2\pi)^4}
\overline{\Psi}^0(p_i) S_0^{-1}(p_i) S(p_i) S_0^{-1}(p_i) \Psi^0(p_i) \right]
(2\pi)^4 \delta^4 \left( \sum_i p_i \right) .
\label{v24}
\end{equation}
The fermion lines pair up into fermion loops due to the color contractions that
appear after the averaging over color orientations. Considering one such loop, we
find
\begin{eqnarray}
I(\rho_1, \rho_2, p, q) & = & - \overline{\Psi}_a^0(p) S_0^{-1}(p) S(p) S_0^{-1}(p)
\Psi_b^0(p) \overline{\Psi}_b^0(q) S_0^{-1}(q) S(q) S_0^{-1}(q) \Psi_a^0(q)
\nonumber \\ & = & {2 (2\pi)^4 \over \mu^2} \rho_1^2 \rho_2^2 { p\cdot q \over p^2
q^2 } f \left( {\rho_1 p \over 2} \right) f \left( {\rho_2 p \over 2} \right) f
\left( {\rho_1 q \over 2} \right) f \left( {\rho_2 q \over 2} \right) + O \left(
\Sigma_0^2 / \mu^2 \right)
\label{v25}
\end{eqnarray}
where in the first line we explicitly show the color indices on the zeromodes. By
defining
\begin{equation}
J(k,\rho_1,\rho_2) = \int I(\rho_1, \rho_2, p, p-k) {d^4p \over (2\pi)^4} ,
\label{v26}
\end{equation}
one can write (\ref{v24}) as
\begin{equation}
G(\rho_1,\rho_2) = \prod_i^{N_f - 1} \left[ \int {d^4k_i \over (2\pi)^4} J(k_i,\rho_1,
\rho_2)  \right] (2\pi)^4 \delta^4 \left( \sum_i k_i \right) .
\label{v27}
\end{equation}
We shall not attempt to evaluate (\ref{v27}) exactly, but instead estimate its size
in terms of the number of fermion loops. It turns out that the $\rho$-integrals are
dominated in the region where $\rho_1\approx\rho_2$. Therefore we set
$\rho_1=\rho_2=\rho$ in the arguments of the $f$-functions. Then (\ref{v26}) becomes
\begin{equation}
J(k,\rho_1,\rho_2) = {8 \pi \rho_1^2 \rho_2^2 \over \mu^2} \int { p^2 - p\cdot k
\over p^2 (p-k)^2 } f^2 \left( {\rho p \over 2} \right) f^2 \left( {\rho
\sqrt{(p-k)^2} \over 2} \right) p^3 \sin^2(\theta)\ d\theta\ dp .
\label{v29}
\end{equation}
For the region where $p>k$ we set $\sqrt{(p-k)^2} = p$ inside the $f$-function and
for $p<k$ we set $\sqrt{(p-k)^2} = k$. Upon evaluating the angular integral we
arrive at
\begin{equation}
J(k,\rho_1,\rho_2) = {(2\pi)^2 \rho_1^2 \rho_2^2 \over \mu^2} \left[ \int_0^k {p^2
\over 2 k^2} f^2 \left( {\rho p \over 2} \right) f^2 \left( {\rho k \over 2} \right)
p\ dp + \int_k^{\infty} \left( 1 - {k^2 \over 2 p^2} \right) f^4 \left( {\rho p \over
2} \right) p\ dp \right] .
\label{v30}
\end{equation}
If we assume that $\rho_1 > \rho_2$ then $\rho = \rho_1$, because the $f$-function
with the largest $\rho$ dominates. Defining
\begin{equation}
x = {p\rho_1 \over 2}  ~~~ {\rm and} ~~~ y = {k\rho_1 \over 2}
\label{v31}
\end{equation}
one can write (\ref{v30}) as
\begin{equation}
J(k,\rho_1,\rho_2) = {\pi^2 \rho_2^2 \over \mu^2} \left[ {8 f^2(y) \over y^2}
\int_0^y f^2(x)\ x^3\ dx + 16 \int_y^{\infty} f^4(x)\ x\ dx - 8 y^2 \int_y^{\infty}
\frac{1}{x} f^4(x)\ dx \right] \equiv {\pi^2 \rho_2^2 \over \mu^2} R(y) .
\label{v32}
\end{equation}
The size of the area under $R(y)$ is $\approx \frac{1}{2}$. This function determines
the momentum dependence under the remaining momentum integrals in (\ref{v27}). To
assess the size of the remaining integrals we note that (\ref{v27}) has $(N_f-2)$
integrals over $k_i$. This will give $(N_f-3)$ angular integrals. (One can arrange
it that each $k_i$ at most contracts with its two neighbors). Together with the
definitions in (\ref{v31}) we estimate the size of the remaining momentum integrals
to be
\begin{equation}
\prod_i^{N_f - 1} \left[ \int {d^4k_i \over (2\pi)^4} R \left( {k_i\rho_1 \over
2} \right) \right] (2\pi)^4 \delta^4 \left( \sum_i k_i \right) \sim \left( {1 \over
2} \right)^{N_f-1} \left[ {16 \over \rho_1^4} (2\pi)^{-2} \right]^{N_f-2} .
\label{v33}
\end{equation}
Our estimate for the sizes of $G(\rho_1,\rho_2)$ then gives
\begin{equation}
G(\rho_1,\rho_2) \sim {\rho_1^4 \pi^2 \over 8} \left[ {2 \rho_2^2 \over \mu^2
\rho_1^4} \right]^{N_f-1} .
\label{v35}
\end{equation}

\section{Results and discussion}
\label{results}

The integrand of (\ref{v16}) is invariant under an interchange of $\rho_1$ and
$\rho_2$. Therefore, one can split the $\rho_2$-integration into two regions, divided
by the value of $\rho_1$. First, we consider only the part where $\rho_2 < \rho_1$.
When we substitute (\ref{v35}) into this part we find
\begin{equation}
W_{\rho_2<\rho_1}^{(2)} = {2^{N_f} \pi^2\over 16} {\cal F} \kappa^2 \mu^2 \int \int
\gamma(\alpha(\rho_1))\ \gamma(\alpha(\rho_2))\ D(\rho_1) D(\rho_2) \rho_1^{3-3 N_f}
\rho_2^{3 N_f-7}\ d\rho_1\ d\rho_2 .
\label{v36}
\end{equation}

One can see that the $\rho_2$-integral is dominated by the region near the
$\rho_1$-boundary. For this reason one can set $\gamma(\alpha(\rho_2)) =
\gamma(\alpha(\rho_1))$. The integrand of the $\rho_1$-integral has its dominant
region around some value $\rho_m$, close to $\mu$. Since we consider cases with
large numbers of flavors, the coupling runs slowly. Therefore one can set
$\gamma(\alpha(\rho_1)) = \gamma(\alpha(\rho_m)) = \gamma_0$. 

Next we use (\ref{v21}) to redefine the variables in (\ref{v36}) and we substitute in
(\ref{v22}). The result is
\begin{eqnarray}
W_{\rho_2<\rho_1}^{(2)} & = & \frac{1}{4} \mu^4 {\cal F} \kappa^2 \pi^2 2^{N_f}
\gamma_0^2 \int_{-\infty}^{\infty} \int_{-\infty}^{\infty} \sigma(\omega_1)
\sigma(\omega_2) F(\omega_1) F(\omega_2) \nonumber \\ & & \times
\int_{-\infty}^{\infty} \int_{-\infty}^{\eta_1} \exp([5-3 N_f+i\omega_1] \eta_1)
\exp([3 N_f-5+i\omega_2] \eta_2)\ d\eta_2\ d\eta_1\ d \omega_1\ d\omega_2 \nonumber
\\ & = & \frac{1}{2} \mu^4 {\cal F} \kappa^2 \pi^3 2^{N_f} \gamma_0^2
\int_{-\infty}^{\infty} \int_{-\infty}^{\infty} \sigma(\omega_1) \sigma(\omega_2)
F(\omega_1)  F(\omega_2) {\delta \left( \omega_1+\omega_2 \right) \over 3 N_f - 5 +
i \omega_2}\ d \omega_1\ d \omega_2
\label{v37}
\end{eqnarray}

Now we add the part where $\rho_2 > \rho_1$. This is the same as (\ref{v37}) but
with $\omega_1$ and $\omega_2$ interchanged. Then we evaluate one of the
$\omega$-integrals. The result is
\begin{eqnarray}
W_{2PI}^{(2)} & = & W_{\rho_2<\rho_1}^{(2)} + W_{\rho_2>\rho_1}^{(2)} \nonumber \\ &
= & \mu^4 {\cal F} \kappa^2 \pi^3 2^{N_f} \gamma_0^2 \int_{-\infty}^{\infty}
\sigma(\omega) \sigma(-\omega) \left[ {(3 N_f - 5) F(\omega) F(-\omega) \over (3 N_f
- 5)^2 + \omega^2} \right]\ d \omega .
\label{v38}
\end{eqnarray}
The maximum value of the part of the integrand in the brackets is at $\omega=0$ with
$F(0)^2 \approx 0.3$. We notice that this term does not have an infrared divergence. 

Comparing the result in (\ref{v38}) with the nondynamical term in (\ref{w2}) one can
see that the potential becomes unstable at the origin when 
\begin{equation}
{\cal F} \kappa^2 \pi^3 2^{N_f} \gamma_0^2 \left[ {F(0)^2 \over 3 N_f - 5} \right] >
{N_c N_f \over 2 \pi} .
\label{v41}
\end{equation}
This can be expressed as 
\begin{equation}
{\cal E} \gamma_0 > 1 ,
\label{v43}
\end{equation}
where
\begin{equation}
{\cal E}(N_c,N_f) = {2 F(0) \exp(5/6) \exp(B N_f - C N_c) \over (N_c-2)!} \sqrt
{2^{N_f+1} (N_f-1)! \over (3 N_f-5) (N_f + N_c - 2)! N_c!}
\label{v44}
\end{equation}
only depends on $N_c$ and $N_f$.

\begin{figure} 
\centerline{\epsfysize = 10 cm \epsfbox{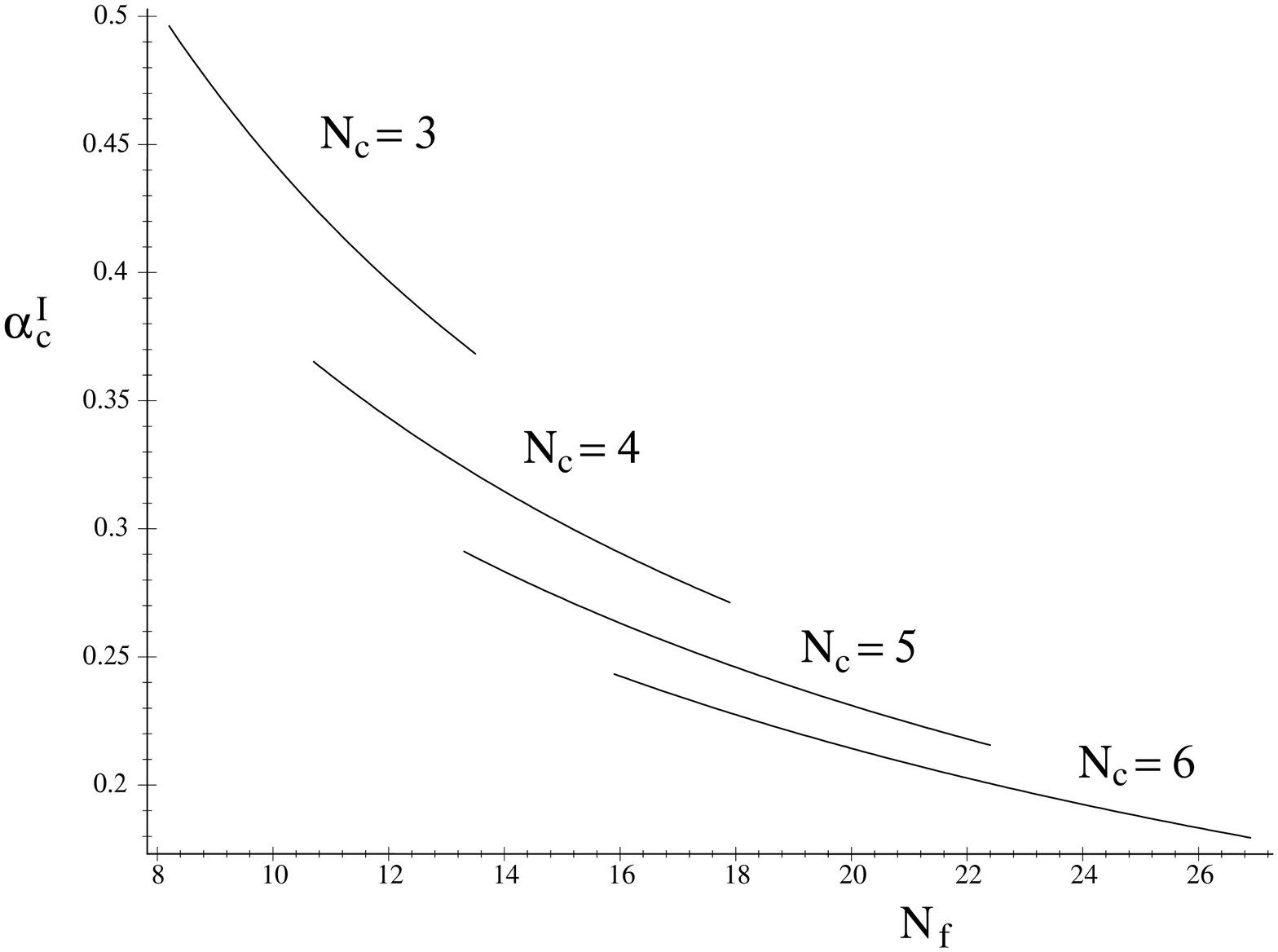}}
\caption{The critical couplings for $N_c = 3, 4, 5$ and $6$, each computed over its
fixed point region up to its critical number of flavors.}
\label{alphac}
\end{figure}

The critical coupling for instanton dynamics $\alpha_c^I$ is the one for which
\begin{equation}
\gamma \left( \alpha(\rho_m) = \alpha_c^I \right) = \frac{1}{{\cal E}(N_c,N_f)} .
\label{v44a}
\end{equation}
It necessarily has to be smaller than $\pi /N_c$. The requirement in (\ref{v44a}),
together with (\ref{v44}), is the main result of our analysis. From it one can
compute $\alpha_c^I$ for any value of $N_c$ and $N_f>4/3$. Numerical estimates of
$\alpha_c^I$ for various values of $N_c$ and $N_f$ are provided in Figure
\ref{alphac}. 

Before we discuss these critical couplings we need to clarify the ranges of numbers
of flavors for which these critical couplings are computed. These ranges are defined
under the assumption that the running of the coupling is governed by the two-loop
beta function -- an assumption which becomes more accurate as the coupling becomes
smaller.  We only compute the critical couplings in the region of $N_f$ where the
two-loop beta function has an infrared fixed point. The left-hand end of each curve
in Figure \ref{alphac} is therefore the point where a fixed point value appears,
which moves down from infinity as a function of increasing $N_f$. 

The right-hand ends of the curves in Figure \ref{alphac} indicate the critical
numbers of flavors $N_f^c$ above which chiral symmetry breaking does not occur.
When there is a nontrivial infrared fixed point in the beta function the maximum
coupling that can be reached is the fixed point coupling. Assuming a two-loop beta
function the fixed point coupling is given by
\begin{equation}
\alpha_* = -\frac{b}{c} = {- 4 \pi (11 N_c - 2 N_f) N_c \over 34 N_c^3 - 13 N_c^2
N_f + 3 N_f}
\label{v45}
\end{equation}
where $b$ and $c$ are respectively the first and second coefficients in the beta
function. The critical numbers of flavors $N_f^c$ is reached when the fixed point
coupling given by (\ref{v45}) falls below the critical coupling, obtained from
solving (\ref{v44a}). Therefore for values of $N_f$ above $N_f^c$ the coupling is
unable to reach the critical value so that chiral symmetry breaking does not occur.
The critical number of flavors, $N_f^c$ are $13.48, 17.95, 22.41$ and $26.87$ for
$N_c=3, 4, 5,$ and $6$, respectively. This gives $N_f^c \approx 4.5 N_c$,
compared to $4.77 N_c$, reported in Reference \cite{r_as}. 

We computed critical couplings for $N_c=3, 4, 5$ and $6$ and for $N_f$ in the fixed
point region up to $N_f^c$. The results are shown in Figure \ref{alphac}. From these
results we make the following observations:
\begin{itemize}
\item The critical coupling decreases as either $N_c$ or $N_f$ increases.
\item The values of the critical couplings near $N_f^c$ are small compared to the
equivalent critical coupling for gauge exchanges given in (\ref{v2}). Compare, for
example, the $N_c = 3$ values. We obtain $\alpha_c^I = 0.37$, while the equivalent
gauge exchange value is $\alpha_c^{ge} = \pi/4 = 0.785$.
\item Toward smaller values of $N_f$ the instanton critical couplings become more
comparable to their gauge exchange counterparts.
\end{itemize}
On this basis we conclude that instanton dynamics are much stronger than gauge
dynamics for large numbers of flavors.

\section{summary}
\label{summary}

We have used the instanton effective action formalism of Reference \cite{r_rt} to
compute the critical couplings for the formation of 2-point functions. For this
analysis we restricted ourselves to continuous phase transitions for chiral symmetry
breaking, which implies that we only consider the instanton--anti-instanton
amplitude with two mass insertions. We found that this amplitude does not suffer
from an infrared divergence. Couplings are allowed to run according to the two-loop
beta functions. The resulting critical couplings indicate that instanton dynamics are
much stronger than gauge dynamics for large numbers of flavors.

\section*{Acknowledgements}

The author wants to express his appreciation for the valuable comments of Bob
Holdom. He also wants to thank Craig Burrell for his help in the preparation of this
manuscript.

\end{document}